\documentclass[onecolumn,showpacs,preprintnumbers,amsmath,amssymb]{revtex4}
\usepackage{amsfonts}
\usepackage{graphicx}% Include figure files
\usepackage{dcolumn}% Align table columns on decimal point
\usepackage{bm}% bold math

\begin{document}

%\preprint{APS/123-QED}

\title{Atomic Frequency standards Based on Pulsed Coherent Light Storage}% Force line breaks with \\

\author{Bo Yan, Yisheng Ma}
 %\altaffiliation{}%Lines break automatically or can be forced with \\
\author{Yuzhu Wang}%
 \email{yzwang@mail.shcnc.ac.cn}
\affiliation{Key Laboratory for Quantum Optics, Center for Cold Atom
Physics, Shanghai Institute of Optics and Fine Mechanics, Chinese
Academy of Sciences, 201800, China
}%

\date{\today}% It is always \today, today,

\begin{abstract}
We propose a new scheme of microwave frequency standards based on
pulsed coherent optical information storage. Unlike the usual
frequency reference where the Ramsey fringe is printed on the
population of a certain state, we print the Ramsey fringe on the
coherence. Then the coherence is detected in the form of a retrieval
light. The central line of the Ramsey fringe can be used as a
frequency reference in an absorption-cell-based atomic frequency
standard. This scheme is free of light shifts as the interrogating
process is separated from the optical pumping processes, and the
cavity pulling effect is negligible due to the low Q requirement.
Encoding the Ramsey interference into the retrieval light pulse has
the merit of high signal to noise ratio and the estimated frequency
stability of shot noise limit is about $2\times10^{-14}$ in 1
second, this scheme is promising for building small, compact and
stable atomic frequency standards.
\end{abstract}

\maketitle

Atomic frequency standards are very stable and accurate frequency
references widely used in high-speed communication and time keeping
system, as well as in fundamental researches\cite{book}. However the
applications demand local oscillators with high frequency stability
in both the short and medium terms. As such, the passive rubidium
frequency standard has received most attention, and that has been
reduced to the smallest size while keeping a relatively good
frequency stability\cite{passive}. But the light shift is still the
major effect limiting the frequency stability to meet the goal. In
order to improve the frequency stability, many new physical concepts
and technologies have been explored, such as electro-magnetic
induced transparency (EIT) frequency standards\cite{ye jun},
coherent population trapping (CPT) masers\cite{CPT maser}, pulsed
optical pumped(POP) frequency standards\cite{light shift}, and so
on. Meanwhile, exciting progresses have been made in the quantum
optical information storage based on
EIT\cite{harris,EITlukin,taiwan}. By manipulating the strongly
coupled excitations of light waves and spin waves slowly propagating
together, one can store quantum information into the media and
retrieve the signal light by switching on the coupling laser after a
short time\cite{nature1,storage review,nature}.

In this Letter, combining the above two domains, we have proposed a
new scheme of microwave frequency standards based on coherent
optical information storage. It provides a new way to improve the
frequency stability of atomic frequency standards. The basic idea of
the pulsed coherent storage (PCS) frequency standards is illustrated
in Fig. 1, and the basic principle can be understood as follows:
first, a signal laser pulse and a coupling laser pulse are applied,
then shutting down the coupling laser which control the group
velocity of coupled excitations (Dark-state Polaritons)\cite{lukin},
and the signal light pulse is coherently converted into a spin wave,
and stored into the media. In this case there is no photon existed,
then manipulate atom states by two microwave pulses which prints the
Ramsey pattern on it. Finally, switching on the coupling laser, the
stored signal light pulse is retrieved, and then detected to obtain
the Ramsey pattern. The central Ramsey fringe can be used as a
frequency reference of cell-atomic frequency standard. The setup of
our proposal shown in Fig. 1 is nearly the same as most groups
working on coherent information storage, such as Lukin's
group\cite{xiao}. The difference is that the cell, which contained
Rb atoms and buffer gas, is placed in a microwave cavity. The cavity
is tuned to the resonant frequency of clock transition of $^{87}$Rb
atoms.

In order to get an analytic formula, we calculate the process based
on a four-level model: as shown in the right of Fig. 2, the level
$|2\rangle$ is the upper level, which can decay to the $|1\rangle$
and $|3\rangle$ levels, but not decay to the level $|4\rangle$. Two
laser beams couple the $|1\rangle$ and $|2\rangle$, $|3\rangle$ and
$|2\rangle$ levels respectively. They are called the signal laser
and the coupling laser. The levels $|3\rangle$ and $|4\rangle$ are
coupled by the microwave. $\Omega_c, \Omega_s$ and $\Omega_m$ are
Rabi frequencies of the coupling laser, the signal laser, and the
microwave excitations. $\gamma_1,\gamma_2,\gamma_3$ and $\gamma_4$
are the relaxation rates which include the effect caused by buffer
gas in the vapor cell. The typical values are
$\gamma_1=\gamma_2=0.5\times10^9s^{-1}$ and
$\gamma_3=\gamma_4=300s^{-1}$. And the decay rate of the upper level
is $\Gamma=\gamma_1+\gamma_2=1\times10^9 s^{-1}$. The assumption of
no decay from $|2\rangle$ to $|4\rangle$ is due to theoretical
self-consistent, and because the decay from $|2\rangle$ to
$|4\rangle$ will not effect the physical result of our theory.

We use the density matrix formalism to deal with this problem, the
motion equation is
\begin{equation}
\dot{\rho}=-\frac{i}{\hbar}[H, \rho]+\ell\rho.
\end{equation}
Within the rotating wave approximation, the Hamiltonian
is\cite{EITlukin}
\begin{equation}
H=-\hbar\left(
\begin{array}{cccc}
\Delta_c-\Delta_s  & \Omega_s & 0 & 0\\
\Omega_s & \Delta_c & \Omega_c & 0\\
0 & \Omega_c & 0& \Omega_m\\
0 & 0 & \Omega_m & -\Delta
\end{array}\right),
\end{equation}
where $\Delta_c=\omega_c-\omega_{23}$,
$\Delta_s=\omega_s-\omega_{21}$ and $\Delta=\omega_m-\omega_{34}$
are the detunings of the coupling laser, the signal laser and the
microwave, respectively.  $\ell\rho$ is the decay matrix which is
determined by all the decay rates shown in Fig. 2.

The timing sequences are shown in the Fig. 1. During the first pulse
$t_p$, only the microwave and the coupling laser are on,
$\Omega_s=0$ and $\Omega_c, \Omega_m\neq0$. This ``state
preparation" stage aims to pump all the atoms to the $|1\rangle$
level, and destroy all kinds of coherences which may cause light
shifts. There are two pumping rates in this process, the rate of
pumping atoms from the $|3\rangle$ level to the $|1\rangle$ level
which is characterized by $\Gamma_p=\Omega_c^2/(2\Gamma)$, and the
rate of transfering the $|4\rangle$ level to the $|3\rangle$ level
which is characterized by the Rabi frequency of microwave
$\Omega_m$. If $\Omega_m>>\Gamma_p$, the atoms transferred from
$|4\rangle$ to $|3\rangle$ will not be pumped to $|1\rangle$ timely.
It experiences a damped Rabi oscillation of coherences and
populations. The decay rate of the whole system is determined by
$\Gamma_p$; And, if $\Gamma_p>>\Omega_m$, once upon the atoms in the
$|3\rangle$ level are pumped to the $|1\rangle$ level at once, the
total decay rate is determined by $\Omega_m$. So the whole decay
rate is
\begin{equation}
\gamma_d\sim min( \Omega_c^2/(2\Gamma),\Omega_m).\label{decay}
\end{equation}
In Fig. 3, We plot the dynamics of populations and coherences when
$\Gamma_p=50000s^{-1}$ and $\Omega_m=10^{5}s^{-1}$. The whole decay
rate of coherence is about $20\mu s$ as predicted by formula
(\ref{decay}). As an example, we choose $t_p=500\mu s$, all
coherences will be destroyed, and the atoms are pumped to a fixed
state,
\begin{equation}
\rho_{11}=1, \ \ \rho_{31}=\rho_{41}=0.
\end{equation}

After this "state preparation" stage, we switch on the coupling
laser and the signal laser. In this way, an EIT state is built.
There will be coherence between $|1\rangle$ and $|3\rangle$. Under
the assumption $\Omega_s<<\Omega_c, \gamma$, to the first order of
$\Omega_s$, we get a stationary solution
\begin{equation}
\rho_{31}^0=-\frac{\Omega_s\Omega_c}{(\gamma+i\Delta_s)[\gamma_3+i(\Delta_s-\Delta_c)]+\Omega_c^2},
\end{equation}
where $\gamma=\gamma_1+\gamma_2$ is the decay rate of the upper
level. In our case, the two-photon resonance condition is fulfilled,
and $\Delta_s\ll\gamma$. When $\Omega_c^2\gg\gamma\gamma_3$, we have
\begin{equation}
\rho_{31}^0=-\frac{\Omega_s}{\Omega_c},\ \ \rho_{41}^0=0. \label{EIT
state}
\end{equation}
It depends only on the Rabi frequency ratio of the signal and the
coupling laser.

After the EIT state has been built up, we turn off the coupling
laser, and the signal light pulse is coherently converted into a
spin wave, and stored into the media. We apply two microwave pulses
to manipulate the coherences. After that, the coherence between the
 $|1\rangle$ level and the $|3\rangle$ level becomes
\begin{widetext}
\begin{equation}
\rho_{31}(T+2t)=e^{i\frac{\Delta}{2}(T+2t)-\gamma_3
T}[e^{-i\frac{\Delta}{2}T}(cos\xi t-i\frac{\Delta}{2\xi}sin\xi
t)^2-e^{i\frac{\Delta}{2}T}(\frac{\Omega_m}{\xi}sin\xi
t)^2]\rho_{31}^0,
\end{equation}
\end{widetext}
where $\xi=\sqrt{(\Delta/2)^2+\Omega_m^2}$. In this way, a Ramsey
pattern is printed on the coherence.

In the final signal detection stage, we apply a strong coupling
laser pulse to get back the signal light pulse again, i.e, the
retrieval of the light pulse stored in the media. The coupling laser
can be applied adiabatically or nonadiabatically in certain
condition\cite{lukin,scully}, and the retrieval light can keep the
spatial shape of $\rho_{31}$.  But different from quantum
information storage,  what we concern more is  the integration of
the light intensity over time rather than the shape of the retrieved
light pulse, because it is related to the probability of the clock
transition. The detected signal is
\begin{equation}
P=P_0 |\rho_{31}(T+2t)|^2/|\rho_{31}^0|^2.
\end{equation}
$P_0e^{-\gamma_3T}$ is the signal detected when microwave pulses are
not applied. It depends to the atom number N, the Rabi frequency of
the coupling laser $\Omega_c$, the way of switching on and off the
coupling laser, and so on, can be
optimized\cite{optimize1,optimize2}. But in the identical experiment
conditions, $P_0$ is a constant. In the vicinity of
$\Delta\approx0$,
\begin{equation}
P=P_0e^{-2\gamma_3T}[1-2cos^2\xi tsin^2\xi t(1+cos\Delta
T)],\label{Ramsey}
\end{equation}
exhibits a Ramsey pattern. We plot a numerical result in Fig. 4 when
$\Omega_m t=\pi/4$ and $\gamma T=400\pi$. The full-width at half
maximum of the central Ramsey fringe is
\begin{equation}
\Delta\omega_{1/2}=\pi/T.
\end{equation}

When the microwave pulse is a $\pi/4$ pulse, i.e, $\Omega_m
t=\pi/4$, the signal has the maximum signal to noise ratio (SNR).
Since the central signal of this Ramsey fringe is zero, it is a
homodyne detection. When $\Omega_mt=\pi/2$, the Ramsey pattern
vanishes as indicated by equation (\ref{Ramsey}). It can be
understood as follow: we detect the mode of $\rho_{31}$, if the
pulses are $\pi/2$ pulses, there will be a Ramsey pattern printed on
the phase of $\rho_{31}$ which can not be reflected in the intensity
detection.
%
%The difference between usual frequency references and our scheme is
%that: Usually, people prepare atoms in a certain state, and use
%double microwave pulses to manipulate the population of this state,
%which print the Ramsey fringe on the population. And finally detect
%the residual population. In our scheme, we prepare atoms in a state
%with coherence, and use double microwave pulses to manipulate the
%coherence, which also print Ramsey fringe on the coherence. Finally
%we detect the coherence in the form of the retrieval light, and also
%get the Ramsey fringe.

In the scheme of PCS frequency standard there is light shift, but it
can be canceled completely. Usually, a pulsed method is used to
eliminate the light Stark shift. But there may still be light shift
due to "phase memory"\cite{light shift}. It is a common problem in
many schemes. If we use the Ramsey pattern as a frequency standard,
the detection should be done repeatedly and changed the detuning of
microwave step by step. After one cycle, the coherence of
$\rho_{41}$ will decay exponentially in ``state preparation" stage.
But if there is some residual coherence of $\rho_{41}^0$ after the
EIT state has been built, it would shift the central frequency of
the Ramsey fringe.  Consider a residual $\rho_{41}$ exits, so the
initiation condition is different from (\ref{EIT state}),
$\rho^0_{41}\neq0$, and the detected signal will be
\begin{equation}
P=P_0|(cos^2\xi t-sin^2\xi te^{i\Delta T})-isin\xi
tcos\xi t(1+e^{i\Delta T})\frac{\rho_{41}^0}{\rho_{31}^0}|^2. \\
\end{equation}
It leads to a frequency shift $\Delta\nu$ of the central Ramsey
fringe ,
\begin{equation}
tan\Delta\nu T=-\frac{1}{sin2\Omega_m t}\frac{2Re(\rho_{31}^{0*}
\rho_{41}^0)}{|\rho_{31}^0|^2-2Im(\rho_{31}^{0*}
\rho_{41}^0)-|\rho_{41}^0|^2}.
\end{equation}
In the "state preparation" stage, the residual coherence of
$\rho_{41}$ decays exponentially by $e^{-\gamma_d t_p}$ where
$\gamma_d$ is determined by formula (2). Since $\gamma_d t_p>>1$,
the residual coherence is very small. Using the data for formula
(3), the frequency shift can be evaluated by $\Delta\nu
<10^{-10}/T$, which can be considered completely negligible.

As in the usual frequency standards, there also exits the cavity
pulling effect\cite{light shift}. When the microwave frequency is
changed, the amplitude will change. If there is a small difference
between the cavity resonance frequency and the atomic transition
frequency, the amplitude varies asymmetrically, which causes a
frequency shift\cite{book}. And during the free decay time T, the
atoms in the cavity would induce a microwave field which effects the
coherence. It causes a shift of the central Ramsey fringe. The shift
can be evaluated of the order\cite{cavity pulling}
\begin{equation}
\Delta\omega\sim\frac{Q}{T}\frac{\Delta\omega_c}{\omega_{34}},
\end{equation}
where $\Delta\omega_c$ is the cavity detuning from $\omega_{34}$.
When a high Q cavity is needed, such as CPT maser, a high cavity is
needed to get a high SNR for microwave detection signal,
$Q\sim10^5$\cite{CPT maser,physics character}, the cavity pulling
effect should be considered. In our case, we do not need a high Q
cavity. $Q$ can be only several hundreds, for example $Q=300$, the
cavity pulling effect is smaller for a factor of $10^3$ than POP
scheme.

The most important advantage of our proposal is the optical
detection. In the POP experiments, microwave signal is observed to
get the Ramsey fringe. But optical detection is much more sensitive
than microwave detection. The energy of an optical quantum wins a
factor of $10^5$ than a microwave quantum. The SNR of optical
detection will be better. So people also want to do optical
detection with absorption signal in POP\cite{optical detection}. But
because the background of absorption signal always exits because the
decay of the population will add a background to signal, a high
contrast signal can not be achieved. In our scheme, the decay of the
coherence do not add a background, so we can get a high contrast
signal.

The contrast C, which is an important parameter in evaluating
frequency stability, is defined as the signal intensity divided by
the background intensity. From equation (\ref{Ramsey}), the contrast
is
\begin{equation}
C=sin^22\Omega_m t.\label{contrast}
\end{equation}
When $\Omega_m t=\pi/4$, it has the maximum contrast $C=1$, when
$\Omega_m t=\pi/2, C=0$, no Ramsey pattern appears as we have
already mentioned. The contrast in our scheme as show in formula
(\ref{contrast}) does not depend on the interrogation time. The
ultimate limit of SNR is the shot noise
\begin{equation}
\frac{S}{N}=\sqrt{N_p C},
\end{equation}
where $N_p$ is the photon number we detect. The frequency stability
of shot noise limit is
\begin{equation}
\sigma(\tau)=\frac{1}{\pi Q_aS/N}\sqrt{\frac{T_{total}}{\tau}},
\end{equation}
where $Q_a=\nu/\nu_{1/2}$ is the quality factor of line shape. For a
long interrogation time T, the quality factor of line shape $Q_a$
become better linear with T, but the SNR become worse with
$e^{-T/(2\tau_d)}$, where $\tau_d=1/\gamma_3$ is the coherence decay
time. A trade-off for high stability is achieved by choosing
$T=\tau_d\sim2\tau_d$. For $N_p=10^9, \nu=6.8G,
\Delta\nu_{1/2}=200Hz, \Omega_mt=\pi/4, C=1, Q_a=3.4\times10^7$,
cycle time $T_{total}=5ms$, we evaluate the frequency stability of
shot noise limit,
\begin{equation}
\sigma(\tau)=2.0\times10^{-14}\tau^{-1/2}.
\end{equation}
It is a very high performance for compact atomic clocks, because the
stability limit is comparable with the typical performance of
H-masers\cite{H maser stability}.

Usually, the laser noise will effect the coherence stored in the
media, the retrieval light intensity, consequently the frequency
stability. Fortunately, in our scheme, the laser noise will not
limit the stability. The coherence, from formula (\ref{EIT state}),
is determined by the Rabi frequency ratio, and is not effected by
the laser amplitude fluctuations because $\Omega_s$ and $\Omega_c$
fluctuate with the same rate as come from one laser beam, and the
ratio is constant. In the detection stage, The retrieval window in
Rb is several hundreds MHz\cite{frequency window}, so the frequency
noise is not a problem. And there is a saturation
effect\cite{saturation}: when the coupling laser is week, the
integration of retrieval light intensity increases as the power of
the coupling laser increases. But when the power of the coupling
laser increases to a certain value, the integration of retrieval
light becomes a constant. So a strong coupling laser will ensure the
amplitude noise not a problem. The frequency stability of our scheme
is much immune to laser noise.

In conclusion, we have proposed  a pulsed frequency standard based
on coherent storage and retrieval. The difference between usual
frequency references and our scheme is that: Usually, one prepare
atoms in a certain state, and use double microwave pulses to
manipulate the population of this state, which print the Ramsey
fringes on the population, and finally detect the residual
population. In our scheme, we prepare atoms in a state with
coherence, and use double microwave pulses to manipulate the
coherence, which also print the Ramsey fringes on the coherence.
Finally we detect the coherence in the form of the retrieval light
and also get the Ramsey fringes. This PCS scheme has several
advantages: a) light shift due to phase memory is eliminated; b) it
only needs a low Q cavity, the cavity pulling effect can be
neglected; c) the detection is a homodyne light detection, making
the signal contrast quite high; d) the stability is quite immune to
the laser noise. This method can also be applied to other elements
and cold atoms. It is very promising to build compact atomic clocks
with high frequency stability.

We acknowledge Tao Hong and Prof. Guoquan Ni for useful discussions
and careful reading and checking of this manuscript. This work was
supported by The Major State Basic Research Development Program of
China under Grant No. 2005CB724507, The National Basic Research
program of China under Grant No. 2006CB921202, The National Natural
Science Foundation of China under Grant No. 10604058.

\begin{figure}
\includegraphics[height=6cm]{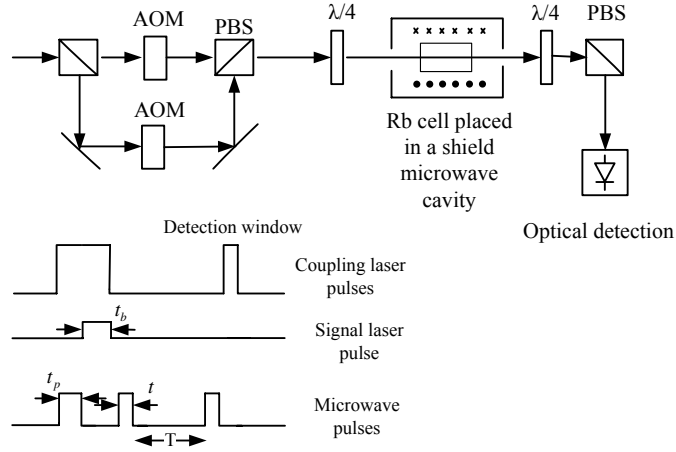}
\caption{The setup of our proposal and the timing sequences of our
scheme. PBS: polarization beam splitter. AOM: acousto-optic
modulator. The Rb cell is placed in a shield microwave cavity. AOM
are used to adjust the two-photon detuning caused by the C-field.
The total cycle time is $T_{total}$, EIT preparation time is $t_b$,
microwave pulse time $t$, and interrogation time $T$. }
\end{figure}

\begin{figure}
\includegraphics[height=4cm]{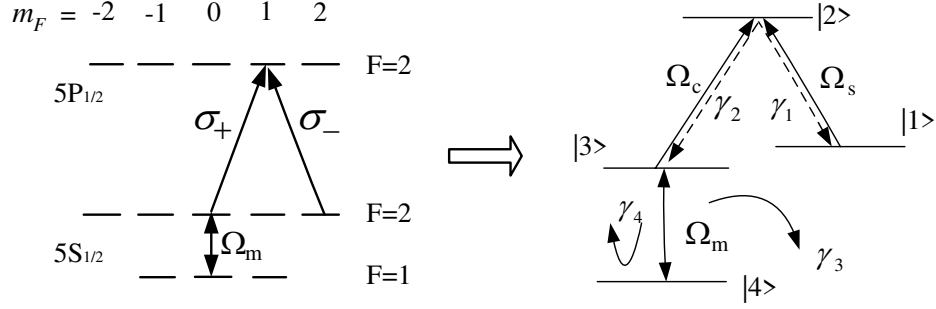}
\caption{The energy levels used in PCS. In the left, it is the
levels used in Rb; In the right, it is the four-level model:
$\gamma_3$ and $\gamma_4$ are the decay rates of hyperfine coherence
and population difference, respectively. $\gamma_1$ and $\gamma_2$
are the decay rates from the level $|2\rangle$ to the level
$|1\rangle$ and the level $|3\rangle$, respectively. We assume there
is no decay from the level $|2\rangle$ to the level $|4\rangle$.}
\end{figure}

\begin{figure}
\includegraphics[height=5cm]{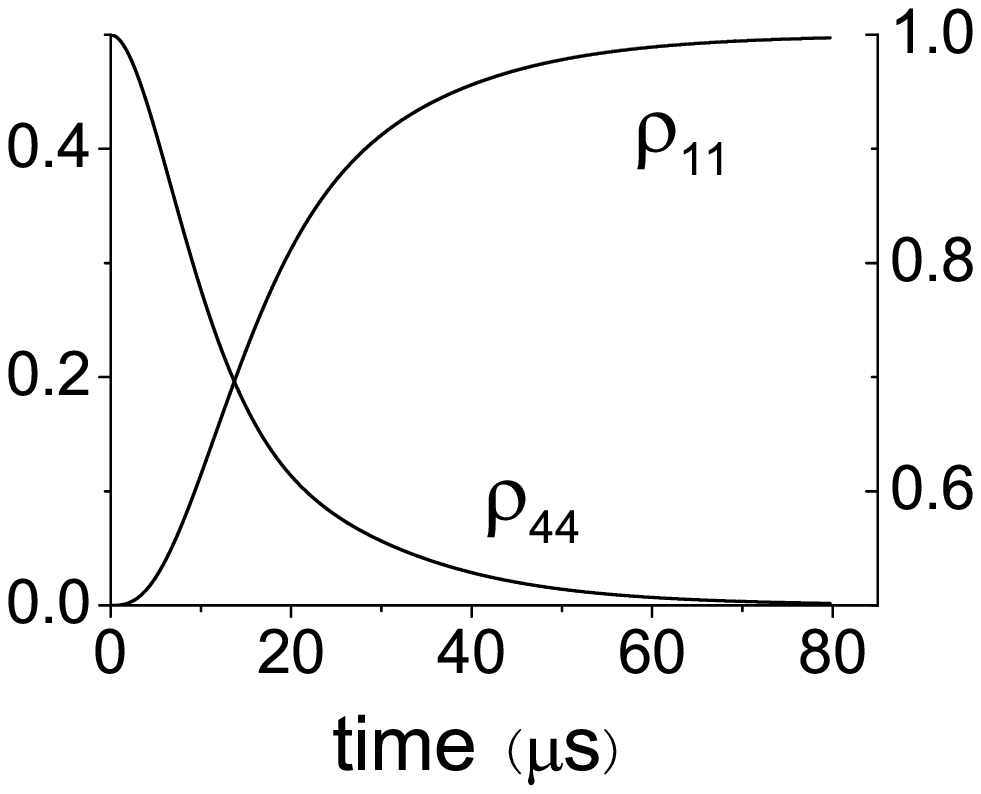}\hspace{1cm}
\includegraphics[height=5cm]{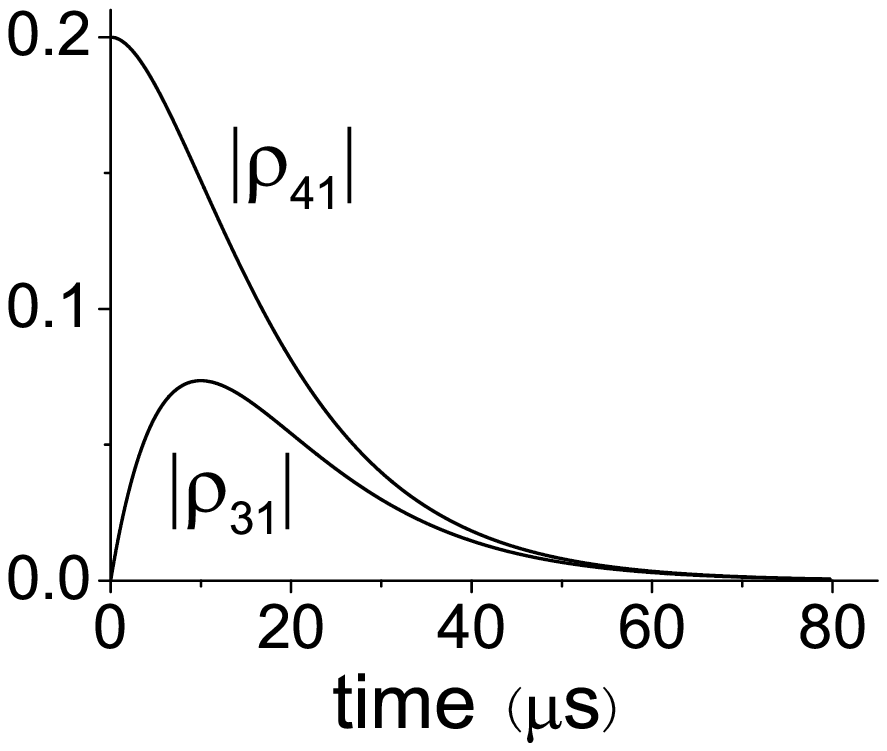}
\caption{In the "state preparation" stage, all atoms will be pumped
to state $|1>$, and destroy all kinds of coherence after a few
hundreds microsecond. We plot the decay of populations and
coherences when $\Gamma=1\times10^9 s^{-1}, \Gamma_p=50000s^{-1}$
and $\Omega_m=10^{5}s^{-1}$.  }
\end{figure}

\begin{figure}
\includegraphics[height=8cm]{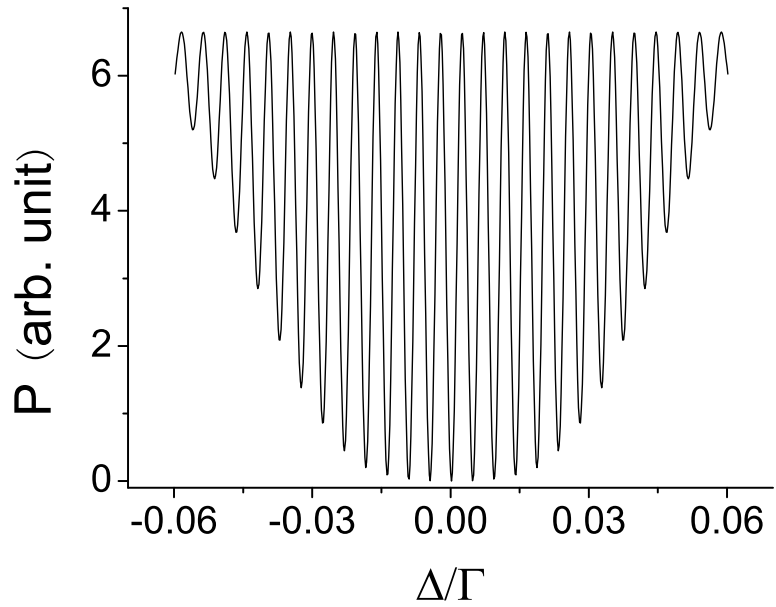}
\caption{The Ramsey pattern of the detected retrieval light when
$\Omega_m t=\pi/4$ and $T=400\pi/\Gamma$. The minimum signal is
zero.}
\end{figure}

\end{document}